\documentclass[aps,twocolumn,superscriptaddress]{revtex4}

\usepackage{graphicx}

\begin{document}

\title{A versatile and compact capacitive dilatometer}

\author{G.~M. Schmiedeshoff, A.~W. Lounsbury, D.~J. Luna, S.~J. Tracy, and A. J. Schramm}

\affiliation{Department of Physics, Occidental College, Los Angeles, California, 90041.}

\author{S.~W. Tozer}
\affiliation{National High Magnetic Field Laboratory, Florida State University, Tallahassee, Florida 32310.}

\author{V.~F. Correa}
\altaffiliation[Permanent address: ]{Comisi\'on Nacional de Energ\'{\i}a At\'omica, Centro At\'omico Bariloche, 8400 S. C. de Bariloche, Argentina}
\affiliation{National High Magnetic Field Laboratory, Florida State University, Tallahassee, Florida 32310.}

\author{S.~T. Hannahs, T.~P. Murphy, E.~C. Palm, and A.~H. Lacerda}
\affiliation{National High Magnetic Field Laboratory, Florida State University, Tallahassee, Florida 32310.}

\author{S.~L. Bud'ko and P.~C. Canfield}
\affiliation{Ames Laboratory and Department of Physics and Astronomy, Iowa State University, Ames, Iowa, 50011.}

\author{J.~L. Smith, J.~C. Lashley and J.~C. Cooley}
\affiliation{Los Alamos National Laboratory, Los Alamos, New Mexico, 87545.}

\date{\today}

\begin{abstract}

We describe the design, construction, calibration, and operation of a relatively simple differential capacitive dilatometer suitable for measurements of thermal expansion and magnetostriction from 300 K to below 1 K with a low-temperature resolution of about 0.05 \AA.  The design is characterized by an open architecture permitting measurements on small samples with a variety of shapes.  Dilatometers of this design have operated successfully with a commercial physical property measurement system, with several types of cryogenic refrigeration systems, in vacuum, in helium exchange gas, and while immersed in liquid helium (magnetostriction only) to temperatures of 30 mK and in magnetic fields to 45 T.

\end{abstract}

%\pacs{}

%\keywords{dilatometry, solids, thermal expansion, magnetostriction}

\maketitle

\section{Introduction \label{Intro}}

It would be difficult to overstate the importance of thermal expansion measurements to the study of solids.  The intimate relationship between thermal expansion and specific heat first explored by Gr\"{u}neisen \cite{Grun1912} has blossomed into a comprehensive theoretical structure \cite{Barron1999} while the characterization of phase transitions using Ehrenfest and Maxwell relations helps coordinate our understanding of the many interesting states exhibited by novel materials.  First-generation samples of such materials are frequently millimeter sized or smaller.  The intense interest in studying these samples at low temperatures (where the thermal expansion can also be small) calls for dilation measurements with sub-angstrom resolution.  Such small length changes represent a significant challenge for the experimentalist.  One of us, (JLS) who likes to say that, `experiments are either easier than they should be or harder than they should be', suggests the de Haas-van Alphen effect and specific heat as examples of the former and latter categories.  In this paper we describe a relatively simple differential capacitive dilatometer which we hope will contribute to the movement of the dilation measurements underlying thermal expansion (and magnetostriction) from the latter to the former category. 

In a capacitive dilatometer the dilation $\Delta{L}$ of a sample of length $L$ manifests as a change in the gap $D$ between a pair of capacitor plates.  For an ideal parallel-plate capacitive dilatometer in vacuum the relationship between the measured capacitance $C$ and $D$ is simply 

\begin{equation}
C = \frac{\epsilon_oA}{D}, \label{eqn:simpleC}
\end{equation}

\medskip\noindent where $\epsilon_o=8.85419$ pF/m is the permittivity of free space and $A$ is the area of the capacitor plates.  Central issues for the researcher include corrections to this simple relationship, the appropriate value of $A$, the temperature and magnetic field dependences of the dilatometer, and any necessary corrections associated with the environment surrounding the dilatometer (liquid helium for example).  Measurements with respect to temperature $T$ yield either the thermal expansivity $\epsilon = (L(T)-L(0))/L(0)$ or the coefficient of linear thermal expansion $\alpha = d\epsilon/dT = (1/L)dL/dT = d(\ln L)/dT$ whereas isothermal measurements with respect to magnetic fields $H$ yield the linear magnetostriction $\lambda = (L(H)-L(0))/L(0)$.  

Researchers considering a capacitive dilatometer design should consult the papers by Pott and Schefzyk \cite{Pott1983}, Swenson \cite{Swenson1998}, and Rotter {\it et al.} \cite{Rotter1998} for recent discussions of the history and capabilities of this approach to dilatometry and for details of capacitive dilatometer design (both in the papers themselves and the extensive references therein).  The principle difference between our design and those described in the references above is in the open architecture of the sample mounting arrangement which permits a wide range of sample shapes and sizes as well as the ability to observe and (to a limited degree) adjust the orientation of the sample in the dilatometer.

In the following sections we discuss the design and construction of the dilatometer, details of its calibration and operation, the corrections we do (and do not) apply to our data, and some measurements on polycrystalline aluminum and nickel samples.  Unless otherwise noted, all experimental results presented in this paper were determined using a copper dilatometer mounted, in vacuum, on the cold-finger of a $^3$He refrigerator.  The capacitance was measured with a digital, self-balancing, three terminal, commercial capacitance bridge \cite{AH_2500A} operating at 1 kHz whose $10^{-7}$ pF resolution corresponds to a dilation limit of about 0.003 \AA\enspace when our dilatometer is operating near 18 pF.  The temperature was determined using commercial resistive thermometers with the manufacturers calibration \cite{Cernox,RuO2}.

\section{The Dilatometer \label{Dilatometer}}

We constructed our dilatometer of oxygen-free high-conductivity (OFHC) copper because of its high thermal conductivity, machinability, relative insensitivity to high magnetic fields, and well known thermal expansion characteristics.  (A titanium dilatometer of the same general design has been constructed for use in very high magnetic fields, and we know of no reason why other materials could not be used instead.)  All copper parts were cleaned with dilute nitric acid (5-10\% by volume in tap water) and annealed at 300 $^o$C for about 3 hours at a pressure of about 6 Pa (using a rotary pump) to reduce internal strains from the machining process.

A schematic of the dilatometer is shown in Fig. 1; the left panel shows a front ``cut-away'' view of the dilatometer in which its components are identified, the right panel represents the dilatometer viewed from the side.  

\begin{figure} 

\includegraphics[width=3in]{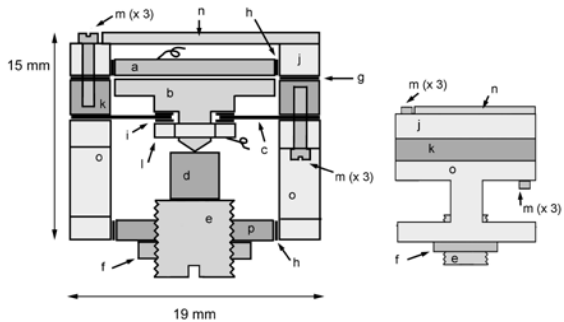}

\caption{A schematic of the capacitive dilatometer.  The left panel shows a front ``cut-away'' view identifying parts: (a) upper (fixed) capacitor plate, (b) lower (movable) capacitor plate, (c) BeCu spring, (d) sample, (e) sample platform, (f) lock-ring, (g) copper shims, (h) electrical isolation (Stycast 2850 FT and kapton), (i) electrical isolation (kapton washers), (j) upper guard ring, (k) lower guard ring, (l) nut, (m) 0-80 copper screws (six in total), (n) mounting plate, (o) main flange, (p) lower flange.  The right panel represents a side view of the dilatometer.  \label{D3fig}}

\end{figure}

\subsection{Assembly Procedure \label{Assembly}}

We discuss the individual parts in the context of their assembly.  While the order of assembly described below need not be rigorously followed some steps must preceed or follow others as noted in the text.  All part designations refer to Fig. 1.

\begin{enumerate}

\item We soldered short manganin wires (represented by the curled lines in Fig. 1) to the upper capacitor plate {\it a} and nut {\it l} to act as electrical contact points for the center conductors of slender, flexible coaxial cables for connection to the capacitance bridge.  We used 60-40 lead-tin solder with manganin wire about 1 cm in length and 0.6 mm in diameter.  The soldering of part {\it a} must preceed its attachment to the upper guard ring {\it j} as described in the following step.

\item The upper capacitor plate {\it a} is attached to the upper guard ring {\it j} with Stycast 2850FT after a thin strip of 25 $\mu$m thick kapton is slipped into the gap between them.  During the gluing process, parts {\it a} and {\it j} are placed on a flat glass plate to ensure their lower surfaces remain coplanar.  Following Swenson \cite{Swenson1997}, the Stycast is allowed to cure for 24 hours at room temperature followed by 24 hours at about 90 $^o$C.  The curing process is followed by a light sanding of the lower surfaces of parts {\it a} and {\it j} to ensure they remain coplanar. 

\item The lower flange {\it p} is attached to the main flange {\it o} using Stycast 2850FT and a thin kapton strip in a similar fashion to the previous step.

\item The spring {\it c}, made of 0.13 mm thick Be-Cu, is bolted to the lower capacitor plate {\it b} between two 25 $\mu$m thick kapton washers using the nut {\it l}.  An aluminum jig is used to hold the spring and lower capacitor plate concentric during this operation.  Three small holes (not shown) penetrate the spring to ventilate the capacitor gap.  We used unannealed commercial ``Alloy 25'' (or C17200); the magnetic susceptibility of this material is somewhat higher than other Be-Cu alloys but the temperature dependence of the susceptibility is smaller \cite{Cooley1995}.

\item The spring {\it c} is positioned between the main flange {\it o} and lower guard ring {\it k} as shown in Fig. 1; the assembly is bolted together with three ``0-80'' OHFC copper screws passing through the main flange and spring before threading into the lower guard ring.  The screws are evenly torqued to about 0.035 N-m (5 inch-ounces).

\item An appropriate number of copper shims {\it g} are positioned between the upper guard ring {\it a} and lower guard ring {\it k} as shown in Fig. 1; the assembly is bolted together with three 0-80 copper screws passing through the upper guard ring and shims before threading into the lower guard ring.  The screws are evenly torqued to about 0.035 N-m.  The number and thickness of the shims are chosen to give a ``zero-force'' capacitance (the capacitance of the dilatometer with no force applied to the lower capacitor plate {\it b}) near 10-13 pF, this value is discussed below.

\item The center conductors of two coaxial cables, eventually leading to the capacitance bridge, are soldered to the two manganin stubs on parts {\it a} and {\it l}.  The shields of the coaxial cables, stripped back from the center conductors, are soldered to a small tab of copper foil; the foil is slipped between the upper guard ring {\it j} and the mounting plate {\it n} to ground the shields at the body of the dilatometer (this step may be modified depending upon the requirements of the capacitance bridge used). The connection to the upper capacitor plate {\it a} passes through a small notch machined into the top of the upper guard ring {\it j}.  The connection to the nut {\it l} passes through a series of concentric holes machined through the upper guard ring {\it j}, the copper shims {\it g}, the lower guard ring {\it k}, the spring {\it c}, and the main flange {\it o}.  The mounting plate {\it n} is bolted to the upper guard ring {\it j} with three 0-80 copper screws.

\item Finally, the sample and the sample platform {\it e} are positioned appropriately (see discussion below) and fixed in place with the lock-ring {\it f}.  We have three sample platforms with incrementally varying lengths to accommodate different sample sizes in different dilatometer mounting orientations.  Our longest sample platform can be secured in place by the lock-ring while pushing the lower capacitor plate in far enough to close the gap (and short the capacitor).  For the adjustments discussed below, it is helpful to use a very fine thread; we are currently using 3.15 threads-per-mm (80 threads-per-inch) but a finer thread would be even better.

\end{enumerate}

The dilatometer may now be tested at room temperature either on the bench or after mounting on an experimental probe or refrigerator via a bolt circle in the mounting plate.  The dilatometer can be mounted in any orientation depending upon experimental requirements (and has operated successfully in a cryostat designed to rotate it {\it in situ}).

\subsection{Testing and Calibration \label{Cal}}

Our task is to find an appropriate functional relationship between the capacitor gap $D$ and the measured capacitance $C$.  Here we describe an approach that may be implemented with the dilatometer either mounted on a cold finger or clamped on a lab bench while using a sample platform {\it e} long enough to adjust the capacitor gap from its largest (zero-force) to its smallest (shorted) value.  

\begin{figure} 

\includegraphics[width=3in]{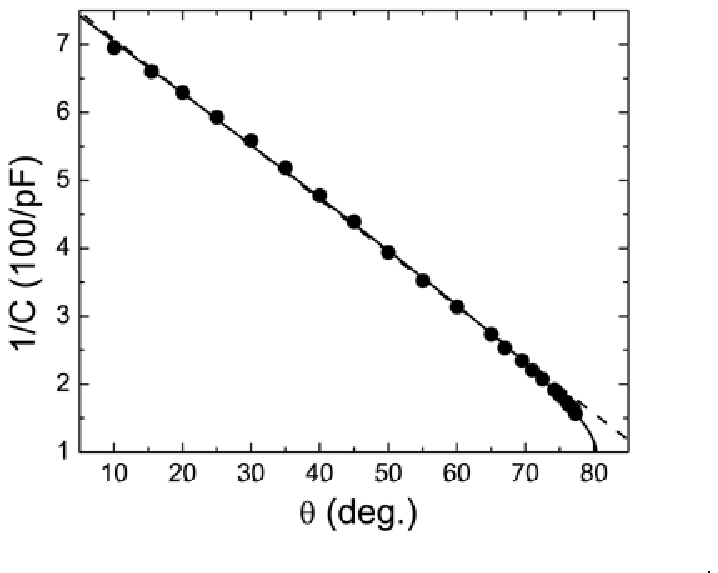}

\caption{Typical dilatometer calibration data:  $\theta$ represents the angular position of the sample platform as it is rotated (decreasing the capacitor gap and increasing the capacitance) in small steps, $C$ is the capacitance after each step. The dashed line is a linear fit to the data with $\theta < 90^o$.  The solid line is a fit to the data incorporating the ``tilt correction'', see text.  Both lines are extrapolated beyond the fit range for clarity. \label{CalFig01}}

\end{figure}

For the calibration data presented below we bolted a dilatometer to a small sheet of aluminum which rested on a flat surface, in air, at room temperature and  attached a protractor (with an appropriately sized hole in its center) to the main flange {\it o}. The dilatometer was inverted with respect to its orientation in Fig. \ref{D3fig}.  The sample platform {\it e} is then rotated (tightened) in small steps; after each step the angular position of the sample platform $\theta$ (read off the protractor), and the capacitance $C$ are measured. The results, plotted as $1/C$ vs. $\theta$ are shown in Fig. \ref{CalFig01}.  The capacitor gap $D$ is related to $\theta$ by 

\begin{equation}
D = c_1(\theta_M - \theta), \label{eqn:D3_cal1}
\end{equation}

\medskip\noindent where $\theta_M$ is the angle at which the dilatometer shorts and the constant $c_1$ is related to the thread pitch on the sample platform.  (For our sample platform with 3.15 threads-per-mm $c_1 = 882.$ nm/deg.)  If the simple parallel plate capacitor model of Eq. \ref{eqn:simpleC} holds then (neglecting the dielectric constant of air)

\begin{equation}
\frac{1}{C} = \frac{c_1}{\epsilon_oA}(\theta_M - \theta). \label{eqn:D3_cal2}
\end{equation}

\medskip\noindent Thus, a plot of $1/C$ vs. $\theta$ should be a straight line whose slope yields the effective area of the capacitor plates (a traditional means of incorporating the edge effects of the capacitive geometry).  

Typical calibration data and a linear fit to the data with $14\enspace{\rm pF} \leq C \leq 43\enspace{\rm pF}$ are shown as the solid symbols and dashed line respectively in Fig. \ref{CalFig01}.  The linear fit yields an effective capacitor area within 1\% of the ``bare'' capacitor plate area measured directly before assembly $A_o = 1.27\times{10}^{-4}$ m$^2$ (the uncertainty in $A_o$ itself is also about 1\%). The uncertainty in the effective area calculated from the fit, incorporating the uncertainties in the fit parameters and $c_1$, is less than 2\%.  This agreement is consistent with estimates of the edge effects expected for our capacitive geometry (two identical, circular capacitor plates separated by a small gap from a grounded, concentric shield) based on exact solutions for a related geometry \cite{Heerens1975} that can be adapted to ours, estimates suggesting that deviations from $A_o$ should be less than 1\%.  Additional corrections associated with the roughness and curvature of the capacitor plates are discussed by Swenson \cite{Swenson1998} and are deemed small enough to ignore.

Note that at high capacitances the data in Fig. \ref{CalFig01} deviate from linearity.  We believe this is due to the fact that the capacitor plates are not perfectly parallel to each other.  Pott and Schefzyk \cite{Pott1983} found an expression for a tilted circular ``parallel'' plate capacitor which Swenson \cite{Swenson1998} expresses as

\begin{equation}
D = \frac{\epsilon_oA}{C}\left[1+\left(\frac{C}{C_{MAX}}\right)^2\right], \label{eqn:tilt}
\end{equation}

\medskip\noindent where $D$ represents the distance from the center of the flat capacitor plate to the center of the tilted plate and $C_{MAX}$ is the capacitance just as the capacitor shorts, a quantity which can be estimated as part of the procedure described above. For the calibration data of Fig. \ref{CalFig01}, $C_{MAX}$ was larger than 65 pF (our bridge overloaded beyond this value, but the capacitor plates did not immediately short).  Disassembling and reassembling the dilatometer, even with the same parts, can significantly affect the value of $C_{MAX}$ (one dilatometer we are currently using, for example, reached 105 pF before the bridge overloaded).  

All of the data shown in Fig. \ref{CalFig01}, expressed as $\theta=\theta(C)$, can be fit to a functional form found by equating Eq. \ref{eqn:D3_cal1} with Eq. \ref{eqn:tilt}.  If the effective area and $C_{MAX}$ are allowed to vary in the fit we find $C_{MAX} = 102$ pF and an effective capacitor area 6\% larger than $A_o$ (larger than the 2\% uncertainty in the calculated effective area).  This fit is represented by the solid line in Fig. \ref{CalFig01}.  However, a fit to the data in which the effective area is fixed and equal to $A_o$ yields $C_{MAX} = 155$ pF which is not physically unreasonable (though the fit is not as good).  For the data on aluminum described below we will use $A_o$ and Eqn. \ref{eqn:simpleC} to convert our measured capacitances to capacitor gaps.  The agreement between our results and those in the literature, discussed in detail below, leads us to suspect we may be accessing a lower bound on $C_{MAX}$ with this calibration procedure and analysis because, for the conditions under which the aluminum data were acquired, Eq. \ref{eqn:tilt} requires $C_{MAX} = 180$ pF for a 1\% deviation in the effective area.  However, after carrying out this calibration procedure and estimating $C_{MAX}$, one could simply choose a capacitance for data acquisition such that Eq. \ref{eqn:tilt} yielded an effective area within 1\% of $A_o$. 

\section{Sample Installation and Data Acquisition}

Generally speaking, to determine either the temperature dependent thermal expansion or the field dependent magnetostriction of a sample, two sets of data are required: one with the sample installed in the dilatometer and one with a known reference material installed in the dilatometer (we use OFHC copper).  An expression relating these two data sets to the thermal expansion of the sample can be derived assuming that the distance between the inner flat surface of the fixed capacitor plate {\it a} and the outer flat surface of the lower flange {\it p} depends upon temperature but is independent of the of the length or nature of the sample installed in the dilatometer.  If the sample is of length $L$ and the reference material is copper (or, more generally, the same material as the dilatometer), then one can show that 

\begin{equation}
\alpha = \frac{1}{L}\frac{dL}{dT} = \frac{1}{L}\frac{d}{dT}\left[D_c - D_s \right] + \alpha_{\rm Cu}\left[1+\frac{D_s-D_c}{L}\right], \label{eqn:cell_effect2}
\end{equation}

\medskip\noindent where $D_c$ is the capacitor gap when the copper standard is mounted in the dilatometer, $D_s$ is the gap when the sample is mounted, and $\alpha_{\rm Cu}$ is the thermal expansion of copper taken from the literature \cite{Kroeger1977}.  First note that this result is independent of the length of the standard material used (which follows from the standard being made of the same material as the dilatometer).  Second, for our cell operating near 18 pF and a sample of length $L = 1$ mm, $D_s/L \sim D_c/L \sim 0.06$ which may not be small enough to ignore, but if the two capacitances remain near 18 pF and differ by 1 pF then $(D_s-D_c)/L \sim 0.003$ which may be small enough to ignore.  If we drop the final term, Eq. \ref{eqn:cell_effect2} can be expressed as

\begin{equation}
\alpha = \frac{1}{L}\frac{dL}{dT} = \left.\frac{1}{L}\frac{dL}{dT}\right|_{\rm Cell + Sample} - \left.\frac{1}{L}\frac{dL}{dT}\right|_{\rm Cell + Cu} + \alpha_{\rm Cu}, \label{eqn:cell_effect1}
\end{equation}

\medskip\noindent since $dL=-dD$.  The first term on the right side of Eq. \ref{eqn:cell_effect1} represents measurements with the sample installed in the dilatometer.  The second term represents measurements with a copper standard installed in the dilatometer, this term is also known as the ``cell effect''.  

Although a conventional OFHC copper sample can be used as a standard, we use the sample platform itself by rotating it until its upper surface is pressed against the rounded point of the lower capacitor plate, the platform is then fixed in position with the lock-ring.  We generally operate with the dilatometer set at about 5 pF greater than the zero-force capacitance (about 18 pF and 13 pF respectively for the dilatometer used for most of the data presented in this paper).    

If the sample has two appropriately positioned parallel faces it can be placed in the center of the sample platform which is then rotated until the sample is pressed against the rounded point of the lower capacitor plate as illustrated in Fig. 1. Once the sample platform is secured with the lock-ring the sample may be gently rotated with tweezers.  We have successfully mounted plate-like samples with only a single flat edge by grasping the sample with tweezers, holding the flat edge against the sample platform, and then rotating the sample platform until the irregular upper surface of the sample is lodged against the rounded point of the lower capacitor plate.  (This sample mounting scheme was devised to accommodate thin, plate-like, rare-earth nickel-borocarbide crystals \cite{Canfield1998,Budko2006a,Budko2006b}, crystals that frequently have irregularities on one or more surfaces.)  

One can use the dilatometer itself to measure the length of irregularly shaped samples after they have been loaded: the sample length $L$ will always be the distance that the sample platform is withdrawn relative to its position for the cell effect measurement. This distance can be measured with calipers or a micrometer.  The precision of this measurement method can, in principle, be improved by incorporating the capacitances of the dilatometer for the two configurations, though we do not think that this approach will always be better than using high quality calipers or micrometers.

The open architecture of the dilatometer means that the sample will be exposed to any black-body radiation illuminating the dilatometer from the side and that a portion of the capacitance circuit, connected to the lower capacitor plate, will not be completely shielded from electrical interference.  We address both of these issues by surrounding the dilatometer (and the cold finger on which it is mounted) with a copper can that acts as both an electrical and a thermal shield.  Thermometers mounted on the cold finger of our $^3$He refrigerator have always agreed with thermometers mounted on the bottom of the dilatometer within experimental resolution if the temperature is not changing too rapidly. 

We generally take isothermal magnetostriction data with the field changing at a maximum rate of 0.4 T/min.  A small  field and sample dependent hysteresis is generally observed in isothermal magnetostriction data (about $\pm 8$ \AA\enspace near 9 T and 10 K when sweeping the field at 0.3 T/min in measurements of the field dependent ``cell effect'', see below).  We attribute the hysteresis to magnetic forces acting on the eddy currents generated in the dilatometer and sample (unless the sample is insulating) by the changing magnetic field.  The magnetic moments associated with these currents change sign as the field increases or decreases.  Forces arise because the field of our superconducting magnet is not perfectly uniform.  The size of the hysteresis is therefore reduced as the field sweep slows; slower field sweeps also reduce eddy current heating in the dilatometer.  Averaging isothermal data taken with the field increasing and decreasing yields a field dependence of the capacitor gap less than 1 \AA\enspace over 9 T, a value consistent with data taken after changing the field and waiting a few seconds for the hysteretic signal to vanish.

For temperatures from about 20 to 300 K we usually take temperature-dependent data with $T$ increasing continuously at a rate of about 0.4 K/min or less. For our system, data taken while warming are generally less noisy than while cooling in this temperature range.  However, one of us (JCL), running an identical dilatometer in a commercial physical property measurement system, finds that data taken while the experimental region is continuously pumped and while cooling at 0.2 K/min are less noisy. The optimal data acquisition protocol in this temperature range can, as one might expect, vary from system to system.  Below about 20 K we keep to this rate (or slower), but the warming and cooling data exhibit comparable levels of noise.  We generally take data while the temperature is varying rather than stabilizing the temperature and then making the measurement because it requires less time to acquire and avoids small displacements that appear occasionally in the raw data.  We believe these displacements are caused by the sudden relaxation of small strains in the dilatometer or sample being studied or from building vibrations, these events manifest as abrupt changes in the size of the capacitor gap $D$ (on the order of 1-10 \AA). Such ``glitches'' are not uncommon \cite{Meingast1990} and do not affect the slope of the nearby data (from which the thermal expansion is determined, see below).  However, should such an event occur between two data points acquired by first stabilizing the temperature and then measuring the capacitance (instead of continuously monitoring the capacitance during the temperature change), the resulting slope would erroneously incorporate the displacement.

The capacitance of the dilatometer is sensitive to thermal gradients: Though constructed of high thermal conductivity copper, the electrical isolation separating parts of the capacitor circuit impedes heat flow.  The effects of such gradients are apparent, for example, when warming or cooling the dilatometer rapidly.  In such circumstances the capacitance data are offset in opposite ``directions'' from data taken either in equilibrium or with the temperature varying very slowly.  The opposite signs of the temperature gradient across the dilatometer when warming or cooling causes an apparent shift in capacitance, qualitatively proportional to both the sign and magnitude of the thermal gradient.  In steady state situations, however, the slopes of capacitance with respect to temperature are identical within experimental uncertainty as long as the warming or cooling is not too rapid.  These effects can be mitigated, somewhat, if the dilatometer is surrounded by helium exchange gas (though the dielectric constant of the exchange gas may have to be incorporated in the analysis).  Operating the dilatometer under liquid helium provides excellent thermal contact to the sample and dramatically reduces the thermal gradients across the dilatometer.  However, the thermal expansion of the liquid helium (either $^3$He, $^4$He, or $^3$He-$^4$He mixtures) dominates the temperature dependence of the capacitance through the dielectric constant of the liquid.  We have not yet made reliable thermal expansion measurements under liquid helium.

To characterize the resolution of the dilatometer under various operational conditions we first measured the dilatometer capacitance repeatedly over a period of about 30 minutes with two identical dilatometers, one held at 5.00 K in a commercial physical property measurement system \cite{PPMS} surrounded by a small amount of helium exchange gas, and another held at 0.300 K in vacuum.  In both cases the capacitance bridge averaging time was set to about 8 s.  After converting the measured capacitances to capacitor gaps in a manner described above we found an rms deviation from the mean capacitor gap of 0.03 \AA\enspace for the former dilatometer and 0.05 \AA\enspace for the latter.  When the latter dilatometer was warming at a rate of 0.4 K/min over a 2 K range near 20 K the rms deviation from a linear fit to the measurements was 0.04 \AA, near 280 K the rms deviation was 0.11 \AA\enspace (we believe this latter uncertainty could be reduced by using a thermometer more appropriate for this temperature range).

A magnetically anisotropic sample in a magnetic field will experience a torque if its magnetic moment is not parallel to the field.  Any motion of the sample in response to this torque will contribute to the measured capacitance change.  We believe we have seen this effect manifest as an irreproducibility in magnetostriction measurements after the sample has been removed from the dilatometer and subsequently reinstalled.  We are exploring the use of thin films of glue or varnish to affix the sample to the sample platform, preventing it from responding to the magnetic torque.  Unfortunately the glue or varnish will also contribute to the thermal expansion, though if it is very thin its affects may be small enough to ignore.  Further work is underway on this issue.

\section{Data Reduction and Results}

Once the capacitance of the dilatometer has been measured as a function of temperature (assuming, for example, that our goal is to determine the thermal expansion through evaluation of Eq. \ref{eqn:cell_effect1}) the next tasks are to convert the measured capacitances $C$ to capacitor gaps $D$ (as described above) and then to evaluate the derivatives of the resulting data with respect to temperature.  

\begin{figure} 

\includegraphics[width=3in]{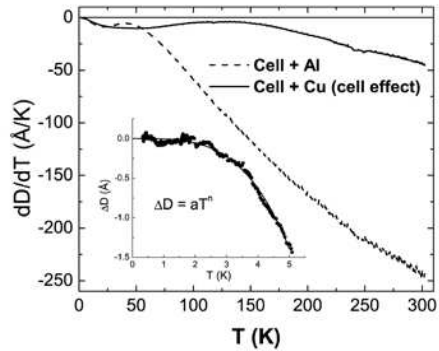}

\caption{Typical data showing $dD/dT$ for the dilatometer with a copper standard (solid line, the ``cell effect'', see text) and with an aluminum sample (dashed line). The inset shows the low-temperature dependence of the capacitor gap for the cell effect (see text); the solid line is a fit to the form $\Delta{D} = D(T) - D(0) = a T^n$. \label{CellFig}}

\end{figure} 

The derivatives required to determine the thermal expansion can be evaluated point-by-point \cite{Barron1998}, or by algebraic polynomial \cite{Barron1980}, cubic spline \cite{Pott1983}, or Chebychev polynomial \cite{Neumeier2005} fits to the data points.  Generally speaking the best approach will depend upon the sample under study.  Problems will arise for any of these methods if the data set contains glitches in $D(T)$ as discussed above.  Cycling the temperature (or the magnetic field) usually reduces the number and size of the glitches which are easy to identify in simple derivatives of the raw data which, for the $i^{th}$ data point we usually evaluate as

\begin{equation}
\left.\frac{dD}{dT}\right|_i = \frac{1}{2}\left[ \frac{D_{i+1} - D_{i}}{T_{i+1} - T_{i}} + \frac{D_{i} - D_{i-1}}{T_{i} - T_{i-1}} \right]. \label{eqn:deriv01}
\end{equation}

\medskip\noindent The glitches in $D(T)$ occur at apparently random temperatures and manifest as delta-function-like features in the derivative; we delete the points forming these features from our data. A numerical integral is performed on the resultant data prior to function fitting (if required).  Typical data, showing the derivatives associated with the first two terms on the right hand side of Eq. \ref{eqn:cell_effect1} (the cell effect and data on an aluminum sample discussed below), are shown in Fig. \ref{CellFig}; these derivatives were evaluated using Eq. \ref{eqn:deriv01}.  The inset shows the temperature dependence of the capacitor gap for the cell-effect below 5 K, the solid line is a fit to the form $\Delta{D} = D(T) - D(0) = a T^n$.  For the data shown the mean deviation from the fit is 0.025 \AA.

Approximately six delta-function-like features were removed from each of the data sets shown in Fig. \ref{CellFig}.  A small feature near 240 K, visible in both data sets, is an artifact of the dilatometer that, to first order, is removed from the final data (discussed below) along with the cell effect.  We have not found examples of cell effects for other dilatometer designs in the literature.  That shown in Fig. \ref{CellFig} is larger than most for the half dozen dilatometers of this design that we have constructed and tested, some of which exhibit a cell effect that changes sign with increasing temperature and one or more features similar to that near 240 K in Fig. \ref{CellFig}.  We continue to explore variations on our annealing and assembly protocols with an eye towards making the cell effect both smaller and smoother.

\begin{figure} 

\includegraphics[width=3in]{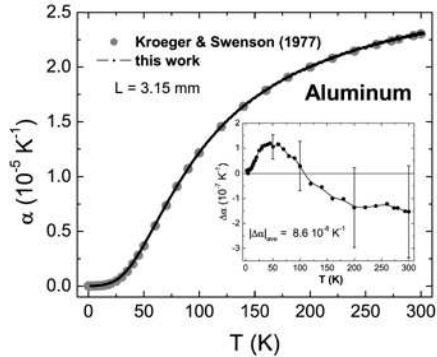}

\caption{The thermal expansion of aluminum, measured by the techniques discussed in this paper (``solid line'', see text), compared to the published values of Kroeger and Swenson \cite{Kroeger1977} (solid circles).  The inset shows the difference between our measurements and tabulated values of Kroeger and Swenson, the solid line is a guide to the eye.  \label{AlFig}}

\end{figure}

Using the techniques discussed above we measured the thermal expansion and magnetostriction of a high-purity sample of polycrystalline aluminum.  The thermal expansion results are shown in Fig. \ref{AlFig} where they are compared to published values.  The solid circles represent tabulated values of the thermal expansion of aluminum \cite{Kroeger1977} from 2 K to 300 K.  The closely spaced results of our measurements, where the derivatives were evaluated using Eq. \ref{eqn:deriv01}, appear to be a solid line.  The inset shows the difference between our measurements and tabulated data in the literature (specifically, Table III in Ref. \cite{Kroeger1977}; in this case the derivatives were evaluated from linear fits to our data in the vicinity of the temperatures of the tabulated results in the literature). The uncertainty bars in the inset are equal to those of our thermal expansion measurements (we take the uncertainty in the literature values to be zero).  The average absolute value of the deviation of our measurements from those in the literature ($\left|\Delta\alpha\right|_{ave}$), over the full temperature range shown, is $8.6\times{10}^{-8}$ K$^{-1}$.  The average absolute value of the fractional deviation ($|\Delta\alpha/\alpha|$) of our aluminum measurements from those in the literature is about 1\% above 40 K but becomes larger at low temperatures where the thermal expansion of aluminum is sensitive to both sample purity and preparation \cite{Kroeger1977}.

\begin{figure} 

\includegraphics[width=3in]{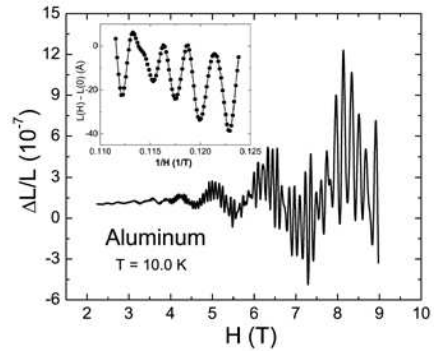}

\caption{The longitudinal magnetostriction of our polycrystalline aluminum sample at 10.0 K showing the characteristic oscillatory behavior resulting from the de Haas-van Alphen effect. The inset shows a portion of the high field data plotted as $\Delta{L} = L(H) - L(0)$ vs. $1/H$. \label{dHvAFig}}

\end{figure}

The longitudinal magnetostriction of our aluminum sample at 10.0 K is shown in Fig. \ref{dHvAFig}.  The field-dependent cell effect for these measurements (analogous to the second term on the right-hand side of Eq. \ref{eqn:cell_effect1}) was less than 1 \AA\enspace over 9 T and is ignored.  The magnetostriction of copper (analogous to the third term on the right-hand side of Eq. \ref{eqn:cell_effect1}) is so small (as expected for a non-magnetic metal) that we are unable to find measurements of it in the literature, making it a natural choice for dilatometers focused on magnetostriction measurements \cite{Fawcett1970}.  The oscillatory magnetostriction, driven by the de Haas-van Alphen effect in the magnetization, is clearly visible.  The dominant oscillation period, about 3.1$\times{10}^{-3}$ T$^{-1}$, is consistent with an ensemble average (our sample is polycrystalline) over the large, high-frequency orbits identified in single crystal measurements \cite{Larson1967}.  The appearance of these oscillations in a polycrystalline sample at such high temperatures suggests that the sample is exceptionally clean. (We have never observed such oscillations associated with the OHFC copper of the dilatometer itself, for example.)

\begin{figure} 

\includegraphics[width=3in]{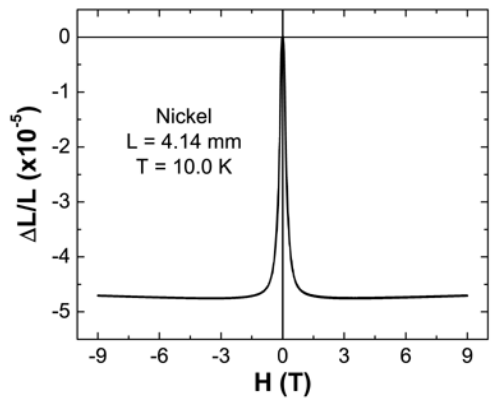}

\caption{The longitudinal magnetostriction of polycrystalline nickel at 10.0 K.  \label{NiFig}}

\end{figure}

\bigskip\noindent The longitudinal magnetostriction of a cylindrical sample of high purity, polycrystalline nickel is shown in Fig. \ref{NiFig}, a pronounced negative Joule magnetostriction (directly related to the field-dependent magnetization) and a slightly positive forced magnetostriction (the magnetostriction in fields beyond that required to saturate the magnetization) are observed.  Both of these features are consistent with published results \cite{Fawcett1967,Nizhankovskii2006}.  

Other measurements of thermal expansion and magnetostriction using dilatometers of this design are beginning to appear in the literature.  Examples include measurements on: the Ising antiferromagnet TbNi$_2$Ge$_2$ \cite{Schmiedeshoff2006}, the non-magnetic borocarbide superconductor YNi$_2$B$_2$C \cite{Budko2006a}, the magnetic borocarbide superconductor ErNi$_2$B$_2$C \cite{Budko2006b}, the shape-memory alloy InTl \cite{Lashley2006a}, the volume collapse in Ce \cite{Lashley2006b}, and the heavy fermion superconductor CeCoIn$_5$ \cite{Correa2006}.  To date the dilatometer has performed well in a range of different laboratory environments and on a diverse set of materials.

The dilatometer, as described, can accommodate sample thicknesses up to about 5 mm, though we know of no reason why the design could not be increased in size to handle larger samples or reduced in size for operation in more confined environments.  The dilatometer has operated successfully with a commercial physical property measurement system (1.8 to 300 K, and in magnetic fields to 14 T), an exchange gas cryostat (1.5 to 300 K), a $^3$He refrigerator with the dilatometer mounted in vacuum (300 mK to 300 K, and in fields to 9 T), as well as $^3$He and dilution refrigerators with the dilatometer immersed in liquid helium (0.25 to about 2 K in fields to 18 T, and 30 mK to about 1.3 K in fields to 45 T respectively, for magnetostriction measurements).  The open architecture, sample mounting scheme, and straightforward relationship between sample dilation and the capacitance of the dilatometer, make this a relatively simple dilatometer to construct and operate.

\section{Acknowledgments}

We gratefully acknowledge the craftsmanship of Roy F. Rockage and helpful conversations with J.~J. Neumeier, M.~F. Hundley, and E.~D. Bauer. Work at Los Alamos National Laboratory was supported under the auspices of the U. S. Department of Energy.  Ames Laboratory is operated for the U. S. Department of Energy by Iowa State University under Contract No. W-7405-Eng.-82. Work at Ames Laboratory was supported by the director for Energy Research, Office of Basic Energy Sciences.  Work at the National High Magnetic Field Laboratory was supported under the auspices of the National Science Foundation, the State of Florida, and the U. S. Department of Energy.  Work at Occidental College was supported by the National Science Foundation under DMR--0305397.

%\bibliography{gms-refs.bib}

\begin{thebibliography}{28}
\expandafter\ifx\csname natexlab\endcsname\relax\def\natexlab#1{#1}\fi
\expandafter\ifx\csname bibnamefont\endcsname\relax
  \def\bibnamefont#1{#1}\fi
\expandafter\ifx\csname bibfnamefont\endcsname\relax
  \def\bibfnamefont#1{#1}\fi
\expandafter\ifx\csname citenamefont\endcsname\relax
  \def\citenamefont#1{#1}\fi
\expandafter\ifx\csname url\endcsname\relax
  \def\url#1{\texttt{#1}}\fi
\expandafter\ifx\csname urlprefix\endcsname\relax\def\urlprefix{URL }\fi
\providecommand{\bibinfo}[2]{#2}
\providecommand{\eprint}[2][]{\url{#2}}

\bibitem[{\citenamefont{Gr{\"u}neisen}(1912)}]{Grun1912}
\bibinfo{author}{\bibfnamefont{E.}~\bibnamefont{Gr{\"u}neisen}},
  \bibinfo{journal}{Ann. Phys.} \textbf{\bibinfo{volume}{39}},
  \bibinfo{pages}{257} (\bibinfo{year}{1912}).

\bibitem[{\citenamefont{Barron and White}(1999)}]{Barron1999}
\bibinfo{author}{\bibfnamefont{T.~H.~K.} \bibnamefont{Barron}}
  \bibnamefont{and} \bibinfo{author}{\bibfnamefont{G.~K.} \bibnamefont{White}},
  \emph{\bibinfo{title}{Heat Capacity and Thermal Expansion at Low
  Temperatures}} (\bibinfo{publisher}{Kulwer Academic/Plenum Publishers},
  \bibinfo{address}{New York}, \bibinfo{year}{1999}), \bibinfo{note}{and
  references therein}.

\bibitem[{\citenamefont{Pott and Schefzyk}(1983)}]{Pott1983}
\bibinfo{author}{\bibfnamefont{R.}~\bibnamefont{Pott}} \bibnamefont{and}
  \bibinfo{author}{\bibfnamefont{R.}~\bibnamefont{Schefzyk}},
  \bibinfo{journal}{J. Phys. E: Sci. Instrum.} \textbf{\bibinfo{volume}{16}},
  \bibinfo{pages}{445} (\bibinfo{year}{1983}).

\bibitem[{\citenamefont{Swenson}(1998)}]{Swenson1998}
\bibinfo{author}{\bibfnamefont{C.~A.} \bibnamefont{Swenson}}, in
  \emph{\bibinfo{booktitle}{Thermal Expansion of Solids}}, edited by
  \bibinfo{editor}{\bibfnamefont{R.~E.} \bibnamefont{Taylor}}
  (\bibinfo{publisher}{ASM International}, \bibinfo{year}{1998}),
  chap.~\bibinfo{chapter}{8}, p. \bibinfo{pages}{207}.

\bibitem[{\citenamefont{Rotter et~al.}(1998)\citenamefont{Rotter, M{\"u}ller,
  Gratz, Doerr, and Loewenhaupt}}]{Rotter1998}
\bibinfo{author}{\bibfnamefont{M.}~\bibnamefont{Rotter}},
  \bibinfo{author}{\bibfnamefont{H.}~\bibnamefont{M{\"u}ller}},
  \bibinfo{author}{\bibfnamefont{E.}~\bibnamefont{Gratz}},
  \bibinfo{author}{\bibfnamefont{M.}~\bibnamefont{Doerr}}, \bibnamefont{and}
  \bibinfo{author}{\bibfnamefont{M.}~\bibnamefont{Loewenhaupt}},
  \bibinfo{journal}{Rev. Sci. Instrum.} \textbf{\bibinfo{volume}{69}},
  \bibinfo{pages}{2742} (\bibinfo{year}{1998}).

\bibitem[{AH_()}]{AH_2500A}
\bibinfo{note}{Model 2500A digital capacitance bridge manufactured by
  Andeen-Hagerling, Inc.}

\bibitem[{Cer()}]{Cernox}
\bibinfo{note}{Cernox thermometer, model CX-1030-SD from Lakeshore Cryotronics,
  Inc.}

\bibitem[{RuO()}]{RuO2}
\bibinfo{note}{Ruthenium Oxide thermometer, model RO-600 from Scientific
  Instruments, Inc.}

\bibitem[{\citenamefont{Swenson}(1997)}]{Swenson1997}
\bibinfo{author}{\bibfnamefont{C.~A.} \bibnamefont{Swenson}},
  \bibinfo{journal}{Rev. Sci. Instrum.} \textbf{\bibinfo{volume}{68}},
  \bibinfo{pages}{1315} (\bibinfo{year}{1997}).

\bibitem[{\citenamefont{Cooley and Aronson}(1995)}]{Cooley1995}
\bibinfo{author}{\bibfnamefont{J.~C.} \bibnamefont{Cooley}} \bibnamefont{and}
  \bibinfo{author}{\bibfnamefont{M.~C.} \bibnamefont{Aronson}},
  \bibinfo{journal}{Journal of Alloys and Compounds}
  \textbf{\bibinfo{volume}{228}}, \bibinfo{pages}{195} (\bibinfo{year}{1995}).

\bibitem[{\citenamefont{Heerens and Vermeulen}(1975)}]{Heerens1975}
\bibinfo{author}{\bibfnamefont{W.~C.} \bibnamefont{Heerens}} \bibnamefont{and}
  \bibinfo{author}{\bibfnamefont{F.~C.} \bibnamefont{Vermeulen}},
  \bibinfo{journal}{J. Appl. Phys.} p. \bibinfo{pages}{2486}
  (\bibinfo{year}{1975}).

\bibitem[{\citenamefont{Kroeger and Swenson}(1977)}]{Kroeger1977}
\bibinfo{author}{\bibfnamefont{F.~R.} \bibnamefont{Kroeger}} \bibnamefont{and}
  \bibinfo{author}{\bibfnamefont{C.~A.} \bibnamefont{Swenson}},
  \bibinfo{journal}{J. Appl. Phys.} \textbf{\bibinfo{volume}{48}},
  \bibinfo{pages}{853} (\bibinfo{year}{1977}).

\bibitem[{\citenamefont{Canfield et~al.}(1998)\citenamefont{Canfield, Gammel,
  and Bishop}}]{Canfield1998}
\bibinfo{author}{\bibfnamefont{P.~C.} \bibnamefont{Canfield}},
  \bibinfo{author}{\bibfnamefont{P.~L.} \bibnamefont{Gammel}},
  \bibnamefont{and} \bibinfo{author}{\bibfnamefont{D.~J.}
  \bibnamefont{Bishop}}, \bibinfo{journal}{Phys. Today}
  \textbf{\bibinfo{volume}{51}}, \bibinfo{pages}{40} (\bibinfo{year}{1998}).

\bibitem[{\citenamefont{Bud'ko et~al.}(2006)\citenamefont{Bud'ko,
  Schmiedeshoff, Lapertot, and Canfield}}]{Budko2006a}
\bibinfo{author}{\bibfnamefont{S.~L.} \bibnamefont{Bud'ko}},
  \bibinfo{author}{\bibfnamefont{G.~M.} \bibnamefont{Schmiedeshoff}},
  \bibinfo{author}{\bibfnamefont{G.}~\bibnamefont{Lapertot}}, \bibnamefont{and}
  \bibinfo{author}{\bibfnamefont{P.~C.} \bibnamefont{Canfield}},
  \bibinfo{journal}{J. Phys. C: Condens. Matt.} \textbf{\bibinfo{volume}{18}},
  \bibinfo{pages}{8353} (\bibinfo{year}{2006}).

\bibitem[{Bud()}]{Budko2006b}
\bibinfo{note}{S.~L. Bud'ko, G.~M. Schmiedeshoff, and P.~C. Canfield, to appear
  in Sol. St. Comm.; cond-mat/0607762}.

\bibitem[{\citenamefont{Meingast et~al.}(1990)\citenamefont{Meingast, Blank,
  B{\"u}rkle, Obst, Wolf, W{\"u}hl, Selvamanickam, and Salama}}]{Meingast1990}
\bibinfo{author}{\bibfnamefont{C.}~\bibnamefont{Meingast}},
  \bibinfo{author}{\bibfnamefont{B.}~\bibnamefont{Blank}},
  \bibinfo{author}{\bibfnamefont{H.}~\bibnamefont{B{\"u}rkle}},
  \bibinfo{author}{\bibfnamefont{B.}~\bibnamefont{Obst}},
  \bibinfo{author}{\bibfnamefont{T.}~\bibnamefont{Wolf}},
  \bibinfo{author}{\bibfnamefont{H.}~\bibnamefont{W{\"u}hl}},
  \bibinfo{author}{\bibfnamefont{V.}~\bibnamefont{Selvamanickam}},
  \bibnamefont{and} \bibinfo{author}{\bibfnamefont{K.}~\bibnamefont{Salama}},
  \bibinfo{journal}{Phys. Rev. B} \textbf{\bibinfo{volume}{41}},
  \bibinfo{pages}{11299} (\bibinfo{year}{1990}).

\bibitem[{PPM()}]{PPMS}
\bibinfo{note}{Physical Property Measurement System manufactured by Quantum
  Design Inc.}

\bibitem[{\citenamefont{Barron}(1998)}]{Barron1998}
\bibinfo{author}{\bibfnamefont{T.~H.~K.} \bibnamefont{Barron}}, in
  \emph{\bibinfo{booktitle}{Thermal Expansion of Solids}}, edited by
  \bibinfo{editor}{\bibfnamefont{R.~E.} \bibnamefont{Taylor}}
  (\bibinfo{publisher}{ASM International}, \bibinfo{year}{1998}),
  chap.~\bibinfo{chapter}{1}, p.~\bibinfo{pages}{1}.

\bibitem[{\citenamefont{Barron et~al.}(1980)\citenamefont{Barron, Collins, and
  White}}]{Barron1980}
\bibinfo{author}{\bibfnamefont{T.~H.~K.} \bibnamefont{Barron}},
  \bibinfo{author}{\bibfnamefont{J.~G.} \bibnamefont{Collins}},
  \bibnamefont{and} \bibinfo{author}{\bibfnamefont{G.~K.} \bibnamefont{White}},
  \bibinfo{journal}{Adv. Phys.} \textbf{\bibinfo{volume}{29}},
  \bibinfo{pages}{609} (\bibinfo{year}{1980}).

\bibitem[{\citenamefont{Neumeier et~al.}(2005)\citenamefont{Neumeier, Tomita,
  Debessai, Schilling, Barnes, Hinks, and Jorgensen}}]{Neumeier2005}
\bibinfo{author}{\bibfnamefont{J.~J.} \bibnamefont{Neumeier}},
  \bibinfo{author}{\bibfnamefont{T.}~\bibnamefont{Tomita}},
  \bibinfo{author}{\bibfnamefont{M.}~\bibnamefont{Debessai}},
  \bibinfo{author}{\bibfnamefont{J.~S.} \bibnamefont{Schilling}},
  \bibinfo{author}{\bibfnamefont{P.~W.} \bibnamefont{Barnes}},
  \bibinfo{author}{\bibfnamefont{D.~G.} \bibnamefont{Hinks}}, \bibnamefont{and}
  \bibinfo{author}{\bibfnamefont{J.~D.} \bibnamefont{Jorgensen}},
  \bibinfo{journal}{Phys. Rev. B} \textbf{\bibinfo{volume}{72}},
  \bibinfo{pages}{220505} (\bibinfo{year}{2005}).

\bibitem[{\citenamefont{Fawcett}(1970)}]{Fawcett1970}
\bibinfo{author}{\bibfnamefont{E.}~\bibnamefont{Fawcett}},
  \bibinfo{journal}{Phys. Rev. B} \textbf{\bibinfo{volume}{2}},
  \bibinfo{pages}{1604} (\bibinfo{year}{1970}).

\bibitem[{\citenamefont{Larson and Gordon}(1967)}]{Larson1967}
\bibinfo{author}{\bibfnamefont{C.~L.} \bibnamefont{Larson}} \bibnamefont{and}
  \bibinfo{author}{\bibfnamefont{W.~L.} \bibnamefont{Gordon}},
  \bibinfo{journal}{Phys. Rev.} \textbf{\bibinfo{volume}{156}},
  \bibinfo{pages}{703} (\bibinfo{year}{1967}).

\bibitem[{\citenamefont{Fawcett and White}(1967)}]{Fawcett1967}
\bibinfo{author}{\bibfnamefont{E.}~\bibnamefont{Fawcett}} \bibnamefont{and}
  \bibinfo{author}{\bibfnamefont{G.~K.} \bibnamefont{White}},
  \bibinfo{journal}{J. Appl. Phys.} \textbf{\bibinfo{volume}{38}},
  \bibinfo{pages}{1320} (\bibinfo{year}{1967}).

\bibitem[{\citenamefont{Nizhankovskii}(2006)}]{Nizhankovskii2006}
\bibinfo{author}{\bibfnamefont{V.~I.} \bibnamefont{Nizhankovskii}},
  \bibinfo{journal}{Eur. Phys. J. B} \textbf{\bibinfo{volume}{53}},
  \bibinfo{pages}{1} (\bibinfo{year}{2006}).

\bibitem[{\citenamefont{Schmiedeshoff et~al.}(2006)\citenamefont{Schmiedeshoff,
  Hollen, Bud'ko, and Canfield}}]{Schmiedeshoff2006}
\bibinfo{author}{\bibfnamefont{G.~M.} \bibnamefont{Schmiedeshoff}},
  \bibinfo{author}{\bibfnamefont{S.}~\bibnamefont{Hollen}},
  \bibinfo{author}{\bibfnamefont{S.~L.} \bibnamefont{Bud'ko}},
  \bibnamefont{and} \bibinfo{author}{\bibfnamefont{P.}~\bibnamefont{Canfield}},
  \bibinfo{journal}{AIP Conf. Proc.} \textbf{\bibinfo{volume}{850}},
  \bibinfo{pages}{1297} (\bibinfo{year}{2006}).

\bibitem[{Las({\natexlab{a}})}]{Lashley2006a}
\bibinfo{note}{J. C. Lashley and R. K. Schulze and B. Mihaila and W. L. Hults
  and J. . Smith and P. S. Risenborough and C. P. Opeil and R. A. Fisher and O.
  Svietelskiy and A. Suslov and A. Planes and L. Manosa and T. R. Finlayson,
  submitted to Phys. B, cond-mat/0607275.}

\bibitem[{Las({\natexlab{b}})}]{Lashley2006b}
\bibinfo{note}{J. C. Lashley and A. C. Lawson and J. C. Cooley and B. Mihaila
  and C. P. Opeil and L. Pham and W. L. Hults and J. L. Smith and G. M.
  Schmiedeshoff and F. R. Drymiotis and G. Chapline and S. Basu and P.S.
  Riseborough, submitted to Phys. Rev. Lett., cond-mat/0608301.}

\bibitem[{Cor()}]{Correa2006}
\bibinfo{note}{V. F. Correa and T. P. Murphy and C. Martin and K. M. Purcell
  and E. C. Palm and S. W. Tozer and G. M. Schmiedeshoff and J. C. Cooley,
  submitted to Phys. Rev. Lett., cond-mat/0609487}.

\end{thebibliography}

\end{document}